\documentclass{aa}
\usepackage[dvips]{graphicx}
\usepackage{lscape}

\usepackage{amssymb}

\begin{document}

   \title{Spectroscopic follow-up of FIRBACK-South bright galaxies}

   \author{J. Patris \inst{1} \fnmsep \inst{2} \thanks{Visiting astronomer, 
European Southern Observatory, La Silla, Chile, under program 66.B-0271} 
 \and M. Dennefeld \inst{1} \fnmsep \inst{2} $^\star$ 
 \and G. Lagache \inst{3}  \and H. Dole \inst{4} 
}
   \offprints{M. Dennefeld: dennefel@iap.fr}
   \institute{Institut d'Astrophysique de Paris (IAP), 98bis Bd Arago,
   	      75014 Paris, France
		\and
	      Universit\'e Pierre et Marie Curie (Paris VI), 4 place
              Jussieu 75005 Paris, France
              	\and
              Institut d'Astrophysique Spatiale, bât. 121, Université
              Paris XI, 91405 Orsay Cedex, France
		\and
	      Steward Observatory, University of Arizona, 933 N Cherry
              Ave Tucson AZ 85721 USA
		}
   \begin{abstract}
   {
	We have performed optical spectroscopy of the  brightest 170$\mu$m 
sources 
   in the FIRBACK South Marano (FSM) field. 
   The spectroscopic sample is $90\%$ complete at the 4$\sigma$ level. 
   Analysis of  spectra and  comparison with spectra of a faint 
   IRAS (60 $\mu$m selected) reference sample is conducted simultaneously. 
   The sources in both samples are characterized by a predominance of 
   emission-line spectra and moderate IR luminosities ($10^{10.5} < 
   L_{IR} < 10^{12} L_\odot$ )
   The fraction of AGN's is low (about $15\%$) and the majority of sources 
   are  nearby ($z<0.3$), dusty,  star forming 
   galaxies, with  moderate star formation rates (a few 10 $M_\odot$
   per year). The infrared emission of the FSM galaxies shows a colder
   spectral energy distribution than was expected.
   The galaxies in both samples (IRAS faint sources and FSM) are essentially
   undistinguishable with the available data, and 
   seem to represent a population of closeby, cold, star forming
   galaxies rather neglected up to now. Although their contribution to
   the far-IR background seems to be low, they deserve a more detailed
   study on their own to assess their importance in the chemical
   evolution of the local universe. The contribution  of fainter,
   presumably more distant, galaxies to the far-IR background will be
   discussed when we  will have completed the follow-up of the fainter part of
   this 170$\mu$m survey.   
   \keywords{galaxies: infrared - galaxies: evolution - cosmology:
   observations }
}
   \end{abstract}
   \maketitle

\section{Introduction}
FIRBACK (Far InfraRed BACKground) is a $170 \mu$m survey carried out by
ISO and designed to study the nature of the CIB (Cosmic Infrared
Background).  It comprises a northern field (FIRBACK N1 and N2; thereafter FN1 and FN2)
and a southern one (the FIRBACK Marano field, FSM). Dole et al. (2001) have 
presented the
global results of the survey and concluded that  a fraction  of the CIB can be 
resolved  into individual sources 
brighter than 135 mJy (the 3 $\sigma$ limit of the survey).
A preliminary analysis by Puget et al. (1999) proposes that  
these sources are a mix of preferentially  
   distant galaxies (at typical
redshifts of 1 to 2), whose detection is favoured in the far-IR by the 
negative K-correction, and which would best reproduce the evolution of the 
number counts (far in excess over those expected from the local FIR population 
discovered by IRAS), and of a  nearby population of very cold galaxies. 
This lead naturally to attempts to observe these sources in the sub-mm range, 
and several 
of them have been detected with SCUBA in the FN1 field (Scott et al. 2000; 
Sajina et al. 2003). 
But until recently, only two redshifts were available, in the FN1 field, two 
optically faint sources ($I=24$ \& 22) 
 observed with the Keck telescope, giving $z=0.5$ and 0.9, respectively 
(Chapman et al.~2002).
These two sources are identified as ULIRG  class galaxies, with merger morphologies
and relatively cold dust temperatures. 
Pending more redshifts, Sajina et al. (2003) used their sub-mm data and 
spectral energy distributions (SED's) to show, 
but on a statistical basis only ,  that FN1 sources appear 
to show a bimodal galaxy distribution, with normal star-forming galaxies 
at z$\simeq$0
and a  z$\sim$0.4-0.9 population  of  much more luminous objects. 
The existence of local, cold galaxies is known from the ISOPHOT serendipity survey 
(Stickel \& al. 2000) which detected a cold ($ \le 20K $) dust component in many galaxies, 
but their  
sample is not complete in any sense and cannot be used for a statistical analysis. 
The FIRBACK sample is therefore the best sample to study the respective importance 
of these two populations and   its  systematic optical follow-up  
remains to be done. \\

 The aim of this paper is to present  a first appraisal of the optical 
counterparts of the  
FIRBACK population (more specifically the brighter sources in the southern 
FSM field) and to derive   their  nature and distance
 by spectroscopy. Spectra in the optical range can provide, 
in addition to redshifts, an analysis of the ionizing source, and an  
estimate of the extinction. 
 In order to compare the FIRBACK objects with well-studied populations
of IR-selected galaxies, we observed also, with the same ESO facilities, 
  a series of  IRAS very
faint sources selected from the sample of Bertin et al. (1997). 
  The IRAS ``Bright Galaxy
Sample'' of Soifer et al (1987), observed in the optical by Kim \&
al. (1995) and analyzed  by Veilleux et al.  
(1995), will also be used as a reference  on bright IR galaxies.

After analyzing all information that can be extracted from the 
spectra, and  comparing these objects with other 
populations, we come to the conclusion that the Marano brightest objects
are part of a very important population of star forming, dusty, closeby  
galaxies, that has been essentially neglected up to now. The question is 
still open for the fainter sources of the FSM survey, which are under  
study  with the ESO VLT. 

The paper is organized as follow: the next three Sections  describe  the 
various data sets and how they were reduced, while  Sect. 5 presents and 
discusses the 
results.  The nature of  FIRBACK bright galaxies is summarized in Sect. 6 
 and the conclusion are presented in the last  Section. 

\section{Data}
\subsection{ISO observations}The FIRBACK survey consists of three separate 
high galactic latitude
regions  which have been observed at 170~$\mu$m using 
the ISOPHOT instrument (Lemke et al., 1996) on board ISO (Kessler et 
al. 1996).
Two of these areas are in the northern sky 
 (FN1 and FN2), while the one we are interested in 
throughout this paper is the southern area, the FIRBACK South Marano or FSM 
field (Dole et al. 2001). 

The FSM region is itself composed of four individual fields,
summing up about 22 sources detected at 4 $\sigma$ and 13 more at a 3
$\sigma$ detection level (Dole et al., 2001), for a total area of 0.95 
square degree. 

 From this source list, we selected   the 
21 brightest objects (FSM\_000 to FSM\_020) for spectroscopic 
follow-up, in order to be complete down 
to the 4 $\sigma$  flux limit  (200 mJy at 170 $\mu$m) in a first phase. 
Three more  objects  at lower IR fluxes (FSM\_026,
FSM\_035, FSM\_039)  have been
observed also during  this run, as a first exploration 
of  the possible nature of fainter sources. 

The coordinates and IR fluxes of those objects are presented in 
Table \ref{list_FIRBACK}. The 
170 $\mu$m fluxes are from the FSM  catalog (Dole et al. 2001) while the 
IRAS fluxes have been obtained from the 
 SCANPI{\footnote{http://irsa.ipac.caltech.edu/applications/Scanpi/}} 
procedure. 

\begin{table*}
 \begin{center}
 \begin{tabular}{lccccc}
  Source name&$\alpha$ 2000&$\delta$ 2000& \multicolumn{3}{c}{Infrared
  flux (mJy)} \\
  &&&ISO 170$\rm \mu$m&IRAS 100$\rm \mu$m &IRAS 60$\rm \mu$m\\ 
  \hline\\
   FSM\_000&$3^{\rm h}09^{\rm{m}}25^{\rm{s}}$&$-54^{\rm o}52'04''$&497$$&570&260 \\ 
   FSM\_001&$3^{\rm h}12^{\rm{m}}07^{\rm{s}}$&$-55^{\rm o}17'09''$&443$$&670&590 \\ 
   FSM\_002&$3^{\rm h}12^{\rm{m}}29^{\rm{s}}$&$-55^{\rm o}16'30''$&420$$&310&170 \\ 
   FSM\_003&$3^{\rm h}11^{\rm{m}}59^{\rm{s}}$&$-55^{\rm o}14'20''$&369$$&$<$219&$<$93 \\ 
   FSM\_004&$3^{\rm h}08^{\rm{m}}37^{\rm{s}}$&$-55^{\rm o}20'45''$&365$$&430&200 \\ 
   FSM\_005&$3^{\rm h}10^{\rm{m}}22^{\rm{s}}$&$-54^{\rm o}31'55''$&301$$&$<$354&$<$81 \\ 
   FSM\_006&$3^{\rm h}10^{\rm{m}}45^{\rm{s}}$&$-54^{\rm o}32'05''$&300$$&$<$186&210 \\ 
   FSM\_007&$3^{\rm h}12^{\rm{m}}10^{\rm{s}}$&$-55^{\rm o}09'00''$&296$$&260&120 \\ 
   FSM\_008&$3^{\rm h}12^{\rm{m}}33^{\rm{s}}$&$-54^{\rm o}57'00''$&269$$&$<$264&$<$120 \\ 
   FSM\_009&$3^{\rm h}08^{\rm{m}}42^{\rm{s}}$&$-54^{\rm o}27'28''$&267$$&370& $<$90 \\ 
   FSM\_010&$3^{\rm h}10^{\rm{m}}16^{\rm{s}}$&$-55^{\rm o}01'37''$&261$$&$<$483&$<$90 \\ 
   FSM\_011&$3^{\rm h}12^{\rm{m}}53^{\rm{s}}$&$-55^{\rm o}09'28''$&239$$&$<$198&$<$78 \\ 
   FSM\_012&$3^{\rm h}08^{\rm{m}}03^{\rm{s}}$&$-54^{\rm o}34'33''$&232$$&$<$306&$<$72 \\ 
   FSM\_013&$3^{\rm h}15^{\rm{m}}18^{\rm{s}}$&$-55^{\rm o}01'26''$&228$$&$<$267&$<$54\\ 
   FSM\_014&$3^{\rm h}14^{\rm{m}}50^{\rm{s}}$&$-54^{\rm o}59'09''$&226$$&$<$159&$<$75 \\ 
   FSM\_015&$3^{\rm h}10^{\rm{m}}37^{\rm{s}}$&$-54^{\rm o}26'16''$&225$$&380&$<$105 \\ 
   FSM\_016&$3^{\rm h}13^{\rm{m}}07^{\rm{s}}$&$-54^{\rm o}49'40''$&214$$&230&$<$72 \\ 
   FSM\_017&$3^{\rm h}08^{\rm{m}}24^{\rm{s}}$&$-54^{\rm o}28'04''$&210$$&-&- \\ 
   FSM\_018&$3^{\rm h}10^{\rm{m}}01^{\rm{s}}$&$-55^{\rm o}11'45''$&207$$&$<$183&$<$57 \\ 
   FSM\_019&$3^{\rm h}07^{\rm{m}}28^{\rm{s}}$&$-55^{\rm o}09'07''$&202$$&-&- \\ 
   FSM\_020&$3^{\rm h}09^{\rm{m}}31^{\rm{s}}$&$-55^{\rm o}25'04''$&200$$&-&- \\ 
  CFSM\_026&$3^{\rm h}09^{\rm{m}}43^{\rm{s}}$&$-54^{\rm o}43'08''$&$160$&-&- \\
  CFSM\_035&$3^{\rm h}08^{\rm{m}}09^{\rm{s}}$&$-55^{\rm o}09'07''$&$142$&-&- \\
  CFSM\_039&$3^{\rm h}10^{\rm{m}}22^{\rm{s}}$&$-54^{\rm o}13'05''$&$134$&-&- \\
  &&&&&\\
 \end{tabular}
 \end{center}
 \caption{List of observed FIRBACK South Marano infrared sources \label{list_FIRBACK}}
\end{table*}

\subsection{IRAS reference sample}

In addition to the FSM objects,  another set of IR selected  
galaxies was analyzed for comparison. These are objects extracted from 
the IRAS Very Faint Source  catalog, in the fields studied 
by Bertin et al. (1997)  and selected 
for their very low 60 
$\mu$m fluxes (in fact the faintest IRAS detections outside the north 
ecliptic pole). The 
sample observed is not  complete in any sense, but contains  a random 
selection  essentially in  the 
 110 mJy~$\lesssim f_\nu(60\mu$ m)$ \lesssim$~200 mJy
 flux range (only a few have stronger fluxes). No particular selection has 
been made, contrary to other samples where only 
ULIRG objects where searched for. 
The technique used to identify those sources is  similar to the 
one used for the FSM sources (see  Sect. 3) and the spectroscopic 
follow-up has been done with the same telescope and similar instruments 
 as for the FSM sources. 
The absolute flux accuracy  of the reduced optical 
spectra in this sample is at the 10\% level. 
The follow-up 
of this reference sample is not complete yet and the final results will be 
published later  (Dennefeld et al., in prep).
Due to their faintness, and to the relative sensitivity 
of the various IRAS channels, in most cases the 100 $\mu$m fluxes are not 
available.

\subsection{Radio data }
A deep continuum 1.4 GHz radio survey (Dole et al.,  in prep.) of
the ISO FIRBACK field has been performed using the
Australia Telescope Compact Array.

A total of 102 hours of observations took place between 
May 14th and 22nd, 1999,
using the 6A configuration, at  1432 MHz and with a
correlator bandwidth of 128 MHz. Twelve pointings cover the ISO field
with  an uniform sensitivity in the center. Data were reduced and
mosaiced using the Miriad package (Sault et al., 1995).

The non-symmetric synthesized beam has a FWHM of approximatively 
$12\arcsec 
\times 10\arcsec$ (in RA, DEC), resulting in a source position accuracy of
at worst $6\arcsec$; with a similar observational strategy, Hopkins et
al. (1998, 2003) and Gruppioni et al. (1999) showed that the
astrometry is better than 2 arcseconds.
The 5 $ \sigma$ sensitivity is at the order of 0.27 mJy.
The Miriad task \verb|SFind|, using the False Discovery Rate method
(Hopkins et al., 2002) was used to extract the sources in the mosaiced
image. \\
The position of the radio sources is marked by dark squares on the 
finding charts (see Fig. \ref{cartes}). Only the positional information 
has been 
used, when necessary, for the FSM source identification, as the flux 
calibration of this radio sample has still to be completed. \\

\subsection{Optical observations} 
The optical follow-up  described here consists  of  
 long slit spectroscopy of the galaxies in 23 fields, corresponding
to essentially to the brighter 
FSM sources. Imaging and photometry of the whole area is also under way, 
but the final data are not yet available. \\

The long slit spectra were obtained on the nights 26-29 November 2000 using 
EFOSC 2 on the 3.6m telescope of ESO in La Silla, Chile. For all 
 spectra, a 1.2$\arcsec$ slit was used with grism  \#13. 
   This setting gives 
a resolution of 2.77\AA\ ~ per pixel and  20\AA\ ~ per slit width, with 
a wavelength coverage from 3680\AA\ to 9340\AA.  It 
allows the identification of many  features 
in the spectra of low $z$ galaxies : in particular the 
$[OII]_{3726-3729}$  emission lines should be detectable for galaxies 
in the 
redshift range $z=0-1.5$, while  the H$\alpha$ Balmer line 
remains in the range for galaxies closer 
than $z=0.4$. The spectral resolution is sufficient  to separate  
H$\alpha$ from 
$[NII]_{6583}$ with the help of fitting procedures. 

The standard procedure with EFOSC2 is  to first take an image of the 
field, and then select interactively the object(s) to be put in the slit. 
 Rotation of the slit allows to obtain at least two 
 spectra in a single exposure if necessary. 
 The acquisition image was obtained with a red Bessel filter (ESO 
 number 642 ), with an exposure time of at least one minute, to allow easy  
 source identification and aperture correction if 
necessary.

The seeing was generally better than  1$\arcsec$ during this observing run 
 and all the  nights were photometric. 

\subsection{Reduction procedure} 
The spectra were reduced and calibrated using ESO's MIDAS 
software in the interactive context ``LONG''. Wavelength 
calibration was performed using Helium Argon lamps spectra, and 
fitting a three degree polynomial. The error on $\lambda$ was found to 
be less than 3 \AA, essentially independent of the
telescope position. 

All the spectra were flux calibrated, after correction of atmospheric 
extinction,
using three spectrophotometric standard stars: LTT 3218, LTT 3864 and 
LTT 1020 (Hamuy et al., 1992 and 1994). The variation between the 
response curves (two observations for LTT 1020, one for LTT 3864 and one 
for LTT 3218)  was found to be about 2\% 
and we used an averaged response to calibrate all scientific 
spectra: we can thus estimate the error on absolute fluxes 
(apart from slit losses)  to be less than 5\%.   

\subsection{Analysis of the  spectra} 

For each calibrated spectrum, the following procedure was employed, making 
use  of ESO/MIDAS routines.

First, a noise spectrum is estimated by summing up the two first 
scales of a wavelet transform  (automatic number of 
scales of 10 for a 2000 pixels spectrum). The two first scales are 
indeed below the resolution of the instrument (20\AA ) and thus can be 
supposed to be pure noise component. This noise spectrum is used: i) 
to estimate the global S/N ratio of the spectrum and ii) to extract  
the high-frequency noise of the whole spectrum and thus make it more 
readable. 

Then the cleaned spectrum is visually examined, and a hypothesis on 
$z$ is made. The hypothesis is tested automatically by searching for 
emission and absorption lines. If more than three separate features 
are found with a $\Delta z$ lower than 0.001, the 
redshift is considered a good guess. 
The program then searches for all lines of a given list. 
Each line found is fitted with a deblend MIDAS routine, with a continuum 
interpolated between two ``flat'' regions (pre-determined, the same for 
all objects) around the line. The total flux, equivalent width and 
measured redshift are given for each line. The final $z$ is determined 
by averaging over all detected features, and the error is given by 
the dispersion - it is generally of the order of $5.10^{-4}$ for 
an average spectrum, and never worse than a few $10^{-3}$.

\begin{landscape} 
\begin{table} 
 \begin{tabular}{ccccccccccccccc} 
   IR source&LIR &ID&z&Av&\multicolumn{2}{c}{$[OII]_{ 
   3727}$}&H$\beta$&$[OIII]_{4959}$ &$[OIII]_{5007}$
   &\multicolumn{2}{c}{H$\alpha$}&$[NII]_{6583}$&Nature&Comments\\
   (1)  &(2)    &(3)&(4)   &(5)&(6) &(7) &(6)  &(6)  &(6)  &(6)  &(7)  &(6)  &(8) &(9)\\
 \hline
FSM\_000& 10.9  &1* &0.0466&3.9&1.4 &256 &0.70 &-    &-    &13   &234  &3.6  &e+a&1 \& 2 interacting\\
	&       &2  &0.0462& - & -  & -  & -   &-    & -   & -   & -   & -   &a   &\\
FSM\_001& 10.6	&1* &0.0297&4.6&0.89&410 & -   & -   &0.14 &5.6  &165  &1.1  &e+a&1 \& 2 interacting\\
	&	&2  &0.0289& - &0.54&26  & -   & -   & -   & -   & -   &0.99 &a   &\\
FSM\_002& 10.4	& * &0.0310&4.3&0.88&269 & -   & -   & -   &4.2  &98   &0.43 &e+a&\\
FSM\_003& 11.8	& * &0.1309&7.1&0.21&2750& -   & -   &0.15 &1.96 &360  &1.1  &e+a&LINER ?\\
FSM\_004& 11.3	& * &0.0782&3.4&4.9 &476 &1.3  &9.2  &24   &11.8 &147  &5.6  &agn &\\
FSM\_005& 9.6	&1* &0.1505& - &0.52&25  & -   &0.21 &0.30 &3.9  &32   &0.81 &e+a&FSM\_005=1+2\\
	&	&2* &0.1516& - &0.19&9.2 & -   &0.22 & -   &1.8  &15   & -   &e+a&1 \& 2 merging\\
FSM\_006&	&1  &0.0480&2.6&0.29&9.1 & -   &0.071&0.13 &0.28 &1.8  & -   &e+a&\\
	& 12.1	&2* &0.2050&2.0&0.68&10  &0.22 &0.087&0.13 &3.5  &15.4 &0.4  &e+a&\\
FSM\_007& 11.2	&*  &0.0923&2.6&1.5 &52  &0.71 &0.29 &0.48 &7.7  &53   &2.3  &e   &\\
FSM\_008& 12.2	&*  &0.2282&2.1&0.34&5.4 &0.18 &0.063&0.23 &2.2  &9.9  &0.92 &e+a&\\
FSM\_009& 11.9	&1* &0.1918&2.0&0.63&9.5 &-    &0.067&0.18 &1.9  &8.3  &0.45 &e+a&interacting with 2\\
	&	&2  &0.1915&2.1&0.23&3.9 & -   & -   & -   &0.56 &2.6  & -   &e+a&\\
FSM\_010& 12.2	&1* &0.2569& - & -  & -  & -   & -   & -   & -   & -   & -   &a   &FSM\_010=1+2\\
	&	&2* &0.2578&1.5&0.25&1.9 &0.082&0.019&0.051&0.72 &2.2  &0.30 &e+a&1 \& 2 interacting?\\
FSM\_011& 12.9	& * &0.6576& - &0.95&45  &0.75 &1.5  &4.7  & -   & -   & -   &agn &pointlike\\
FSM\_012& 12.1  &1* &0.1925&2.5&1.2 &33  &0.43 &0.12 &0.36 &5.0  &31   &1.6  &e+a&\\
	&	&2  &0.0547&4.1&0.79&196 & -   & -   & -   &1.4  &29   &0.15 &e+a&\\
FSM\_013&	&1  &0.0546&1.8&0.40&4.6 & -   & -   &0.15 &0.51 &1.97 &0.030&e+a&\\
	& 12.2	&2* &0.2389&2.2&0.16&3.1 & -   & -   &0.09 &0.46 &2.3  &0.085&e+a&FSM\_013=2+3\\
	&       &3* &0.2541&2.7&0.17&6.5 &0.077&0.14 &0.16 &0.43 &3.2  &0.61 &e+a&LINER ?\\
FSM\_014&11.5(A)&1* &0.1579& - &0.49&23  &0.092&0.071&0.16 &0.27 &2.3  & -   &e+a&IR flux shared A=1+2\\
	&	&2* &0.1589& - &0.61&29  &0.22 &0.10 &0.22 &0.43 &3.6  & -   &e+a&1 \& 2 interacting\\
	&11.8(B)&3* &0.2915&1.6&0.23&2.1 & -   & -   & -   &0.48 &1.6  &0.28 &e+a&IR flux shared B=3\\
FSM\_015&10.8	& * &0.0578&3.9&0.31&56  & -   & -   & -   &0.76 &13   &0.12 &e+a&\\
FSM\_016&11.5	& * &0.1302&3.3&0.50&43  &0.53 &0.079&0.15 &8.3  &96   &0.79 &e+a&\\
FSM\_017&11.0	&1* &0.0744&5.4&0.47&685 & -   & -   &0.091&1.9  &104  & -   &e+a&\\
	&	&2  &0.1295&2.7&0.22&8.2 &0.053& -   &0.081&0.31 &2.3  & -   &e+a&\\
	&	&3  &0.2991& - &0.17&8.2 &0.089& -   & -   &0.17 &1.5  & -   &e+a&\\
FSM\_018&12.0	&1  &0.2329& - & -  & -  & -   & -   & -   & -   & -   & -   &a   &\\
	&	&2  &0.2341& - & -  & -  & -   & -   & -   & -   & -   & -   &a   &\\
FSM\_019&	&1  &0.0725&2.7&0.65&26  &0.084&0.14 &0.084&3.6  &27   & -   &e+a&\\ 
	&12.3	&2* &0.2568&2.2&0.13&2.7 & -   & -   & -   &0.75 &3.9  & -   &e+a&\\
FSM\_020&11.0	&1* &0.0714&2.5&1.2 &35  & -   &0.29 &0.27 &5.3  &33   &0.93 &e+a&1 \& 2 interacting\\
	&	& 2 &0.0713&1.3&1.5 &8.7 &0.32 &0.35 &1.0  &2.0  &5.2  & -   &e+a&\\
CFSM\_026&10.9	&*  &0.0780&4.7& -  & -  & -   & -   & -   &2.0  &66   & -   &e+a&\\
CFSM\_035&12.5	&*  &0.5902& - &0.49&23  &4.1  & -   &1.6  & -   & -   & -   &agn &pointlike\\
CFSM\_039&10.9(A)&1*&0.1413&1.3&1.4 &8.0 &0.38 &0.36 &0.88 &2.8  &7.2  &0.67 &e+a&IR flux shared A=1\\
	&11.7(B)&2* &0.2684&2.2&0.18&3.7 &0.098& -   & -   &1.2  &6.5  &0.44 &e+a&IR flux shared B=2\\ \hline
 \end{tabular} 

 \begin{tabular}{ll} 
(1) Name of FSM 170 $\mu$m published source.&(2) log10 of total IR luminosity 
   in $L_\odot$ (see section 4.5).\\ 
(3) Identification number of the optical object. *: object considered to be 
   the IR source&(4) Redshift measured 
   from lines, error $\simeq 5.10^{-4}$ (see section 2.4)\\ 
(5) Extinction $A_V = R\times E(B-V)$, from 
   Balmer decrement (section 3.3), with $R = 3.2$ & 
(6) Main emission lines fluxes,
$10^{-15} erg/s/cm^2$, error $<$ 10\% \\
(7) Flux corrected from absorption by star \& dust
extinction, $10^{-15} erg/s/cm^2$ (section 3.3) & (8) Spectral
type. $e+a$: emission and absorption lines. $a$: no emission line\\
\multicolumn{2}{l}{(9) Comments. IR flux can be attributed to two
optical objects, see source identification (section 2.5)}\\

 \end{tabular} 	
 \caption{Flux measurements for the FIRBACK South Marano sample. 
\label{results_M}} 
\end{table} 
\end{landscape}  

A rough  spectral  classification (reported
in column 8 of Table \ref{results_M}) is done according to the following 
rules:\\
- $a$ when no emission line has been detected, so that the only
measurable features are the H and K Ca II lines (at 3933\AA\ and
3968\AA) and possibly other absorption lines. \\
- $e+a$ when emission lines have been detected  (H$\beta$ equivalent 
width greater than one \AA, other
non-Balmer emission lines), along with absorption features. \\
- $agn$ when high excitation forbidden lines indicate an active nucleus.

\section{Sources identification} 
Identification of optical  counterparts of far IR sources is a complicated 
process due to the large "beam" size of  ISOPHOT. 
 FSM 170 $\mu$m PHOT sources are ascribed to a 
100 \arcsec diameter circle, which corresponds to a localising probability of 
 about 93\%, as discussed by  Dole et al. (2001). 
The consequence is that usually  more than one  
optical galaxy lies in the error circle, and this multiplicity of course 
 increases when going to fainter optical magnitudes. As we deal here with 
the brightest PHOT sources only, the problem is however less severe, 
and we have, whenever possible, taken spectra of all the ( bright) 
galaxies lying within the error circle. \\

In about half the cases, only one optical galaxy was found (although sometimes 
more than one spectrum was taken, to check other optical candidates, which 
then proved to be stars). If more than one optical galaxy is found inside 
the error circle, we  then used the following criteria to pursue the 
identification further:  \\
- the spectrum should look similar to a "standard"  starburst galaxy, i.e. have 
emission lines (particularly [OII] or H$\alpha$), a reddened continuum, etc.. \\
-  the   radio emission in the ATCA  data at 1.4 GHz (Dole et  
al., in preparation), if available in the field, should correspond to this 
galaxy.  The  better angular resolution (at the 
arc-second level) of the radio  
observations allows in most  cases to identify without ambiguity the 
optical candidate associated with the radio source. 
This general procedure relies on the assumption that the IR PHOT source 
is of a similar nature 
than  the standard IRAS galaxies, powered by star-formation or an AGN, and 
where the  
well  known radio/IR correlation holds over a large range of fluxes. This  
"bias" would, at this early stage, prevent identification of sources of  
different nature, should they exist in this sample. However, the 
``radio emission'' criterion was decisive only for a few objects (four IR 
sources) and we have found no case (in this bright sample) where there  
was only one radio source, corresponding to a non-emission lines galaxy.\\ 
If several optical objects have similar  properties (same redshift, 
emission lines, radio emission) and all lie inside the error circle, 
we assume they  
all contribute to the IR emission and thus we sum all the optical 
contributions. The redshift attributed to the IR object is then the   
average of the optical redshifts . This is the case for four objects: 
FSM\_005, FSM\_010, 
FSM\_013 and FSM\_018. \\ 
More difficult  is the case when two galaxies with different redshifts  
but otherwise similar properties (emission lines, radio emission ...) lie 
in the error circle. Both candidates probably  contribute to the final 
infrared flux, and, since there is  no perfect criterion to chose 
one  candidate or the other, we chose to distribute the IR flux 
proportionnaly to 
the H$\alpha$ line intensity (this assumes a linear relation  
between IR and emission line  fluxes, which holds only roughly). 
These problematic  
cases are fortunately rare enough here (1 case out of 28 for the IRAS  
sample, 2 out of 24 for the FSM sample) that they will not affect our 
general conclusions . We found only one case where none of this criteria 
could be used: in FSM\_018, there is no detected radio emission, and none 
of the observed galaxies does show emission lines (see discussion of individual 
sources).  \\  
\begin{figure}[t] 
\caption{ R-band image of FSM\_005 infrared source, showing galaxy number 
1 (left) \& 2 \label{im5}, taken with theESO  3.6m telescope in La 
Silla.   This 
picture shows clearly the two merging galaxies. The field is rotated 
(the star is due north of 
the galaxies)  in order to place the slit across both 
galaxies). } 
\end{figure} 
Fig. \ref{cartes}  presents the DSS images with a superposed 
infrared error circle, radio identifications (when available),  
 spectrograph slit position and final identification(s) 
(the optical galaxy finally associated with the IR emission is 
indicated by an asterisk, see also Table \ref{results_M}). In the 
following, we discuss each identification separatly.
{\bf
\par\medskip\noindent
\it Main catalog:
\par\medskip\noindent
} 
{\bf$\bullet$ FSM\_000:} Two large galaxies have been placed on the
slit. The infrared source is associated with the
large galaxy inside the error circle, witch is also a radio source. 
 The second galaxy (also a radio source, but outside the 
circle) has not a  starburst spectrum and is therefore assumed to 
contribute little or nothing to the IR emission. \\
{\bf $\bullet$ FSM\_001:} The infrared source is associated with the
large galaxy inside the error circle,  which is also a radio source,
object \#2 being out of the error circle and with no radio emission.\\
{\bf $\bullet$ FSM\_002:} Only one large galaxy is in the error circle,
but there is no radio emission.\\
{\bf $\bullet$ FSM\_003:} Among the numerous stars of the  nearby globular
cluster lies  a  large galaxy, which is also a radio source.\\
{\bf $\bullet$ FSM\_004:} One large radio bright galaxy.\\
{\bf $\bullet$ FSM\_005:} On close examination (see Fig. \ref{im5}),
the central optical source is in fact a couple of  merging galaxies (of whom
one at least is a radio source). We associated the infrared flux with
both optical galaxies (and thus added the emission line intensities).\\
{\bf $\bullet$ FSM\_006:} The galaxy associated with the infrared flux 
(\# 2) is inside the error  circle, and is associated with  radio 
emission. A contribution from galaxy \#1 is possible too. \\
{\bf $\bullet$ FSM\_007:} This large galaxy, 
close to the globular cluster,  is also a radio source.\\
{\bf $\bullet$ FSM\_008:} One galaxy is in the error circle, which is
also a radio source.\\
{\bf $\bullet$ FSM\_009:} There are two galaxies but galaxy \#2 is on the
border of the error circle, has weak emission lines and  no radio 
counterpart. We thus attribute the IR  flux to galaxy~\#1 only.\\
{\bf $\bullet$ FSM\_010:} Two faint galaxies have been observed
spectroscopically. One is well inside the error circle but has no
radio counterpart, the other is on the limit of the error circle and
is associated with a radio source. Since both galaxies are at the same
redshift (though separated by about 250 kpc with a standard, H$_0$=65
km/h/Mpc cosmology) we added the optical fluxes. A much fainter object (not 
observed spectroscopically) and coinciding with a radio source could 
contribute also. \\
{\bf $\bullet$ FSM\_011:} No bright galaxy was  visible inside the 
error circle, so we placed two slits to check various faint  optical 
sources. We found a few stars, two very weak galaxies with no emission 
line, and a point-like active galaxy with bright emission lines and 
rather high redshift ($z=0.66$). The infrared flux was attributed to this 
object. Note that this object has the highest IR luminosity 
of the whole sample, by far.\\ 
{\bf $\bullet$ FSM\_012:} One of the few cases where radio emission 
was decisive: galaxy \#1 was chosen because of it (although galaxy \#2 has 
week emission too).  \\ 
{\bf $\bullet$ FSM\_013:} Two of the three galaxies in the slit are 
well inside the error circle, with similar redshifts. They are 
probably interacting, and thus we considered that both contributed to
the infrared flux (optical fluxes were added).\\
{\bf $\bullet$ FSM\_014:} All galaxies lie within the error circle,
with no radio identification available. Without further information 
available, 
 we divided the infrared flux in two parts, proportionnally to the
H$\alpha$ emission line strength: one part (infrared object FSM\_014 A)
is attributed to galaxies \# 1 and \#2 (two galaxies probably
interacting at $z=0.16$, optical fluxes added), and the other part
(infrared object FSM\_014 B) to  galaxy \#3
($z=0.3$).\\ 
{\bf $\bullet$ FSM\_015:} One large galaxy at the center of the error
circle, no radio emission.\\
{\bf $\bullet$ FSM\_016:} One large galaxy  at the error
circle's center, with radio emission.\\ 
{\bf $\bullet$ FSM\_017:} Two galaxies are inside the error circle, we
chose \#1 because it has  radio  emission.\\
{\bf $\bullet$ FSM\_018:} We observed spectroscopically both  
galaxies inside the error circle, however, none of them shows any 
emission line. There is no radio emission. This would be  the only case where 
the optical counterpart has no emission line. The identification here is 
still unsecure. \\
{\bf $\bullet$ FSM\_019:} Only one galaxy lies inside the error
circle, has emission lines but no radio emission: we consider it as the IR 
counterpart. Note however that another faint source lies on the borderline, 
with radio emission (which was not known at the time of our observations), 
and could contribute to the IR flux also. \\
{\bf $\bullet$ FSM\_020:} One large galaxy inside the error
circle, with optical emission and radio emission.
  
{\bf
\par\medskip\noindent
{\it Complementary catalog:}
\par\medskip\noindent
}
{\bf $\bullet$ FSM\_026:} One large galaxy at the center of the error 
circle. Weak optical emission but No radio emission.\\ 
{\bf $\bullet$ FSM\_035:} From the two radio sources, one seems to be  a quiet 
galaxy (no redshift determined) and the other a bright point-like radio-loud 
QSO at $z=0.6$. We thus  
provisionnally attribute the infrared flux to the AGN alone. \\ 
{\bf $\bullet$ FSM\_039:} Two galaxies with no radio detection lie in 
the error circle, \#1 at $z=0.14$, rather bright but off-centered, \#2 
at $z=0.27$, fainter, centered. We divided the IR flux proportionnally 
to the H$\alpha$ emission line intensity.\\ 

\section{Various corrections} 
\subsection{Aperture correction} 
We have used a  1.2$\arcsec$ slit for the observations, wider than the seeing 
during the whole run. While for point-like source, we may then 
consider that most of the object's flux is transmitted, 
for extended galaxies, on the contrary, a fraction of the flux is lost at 
the entrance, and should be evaluated .
The question of aperture correction  is however difficult, and 
different solutions have been used in the literature:\\
- correction using photometry: the total 
spectrophotometric flux can be corrected by comparing it to the 
photometric flux (magnitude)
 (solution used for instance in the CFRS, 
Tresse et al. 1996).  However, this procedure assumes a regular 
distribution of the surface brightness over the galaxy, 
 and when emission lines are concerned 
 (our main interest here)  is far from being satisfactory. If (as is 
probably the case for our objects)
 the star forming region is essentially circumnuclear, 
 such a correction would grossly   overestimate 
the total emission.\\
- no correction at all, if the slit is considered wide enough 
 (e.g. in Sullivan et al, 2000 for the 
FOCA UV-galaxies follow-up).\\
Our case is intermediate between these two extremes, but because the 
emission is mostly circumnuclear in IRAS galaxies,  we decided not 
to correct the spectra for this effect. In doing this, the measured 
 fluxes are certainly lower limits, but should not greatly 
underestimate the real values for absolute fluxes. It does not affect, 
of course, equivalent widths and flux ratios.  

\subsection{Underlying stellar absorption lines} 
\begin{figure}
\includegraphics[width=5.5cm,angle=270]{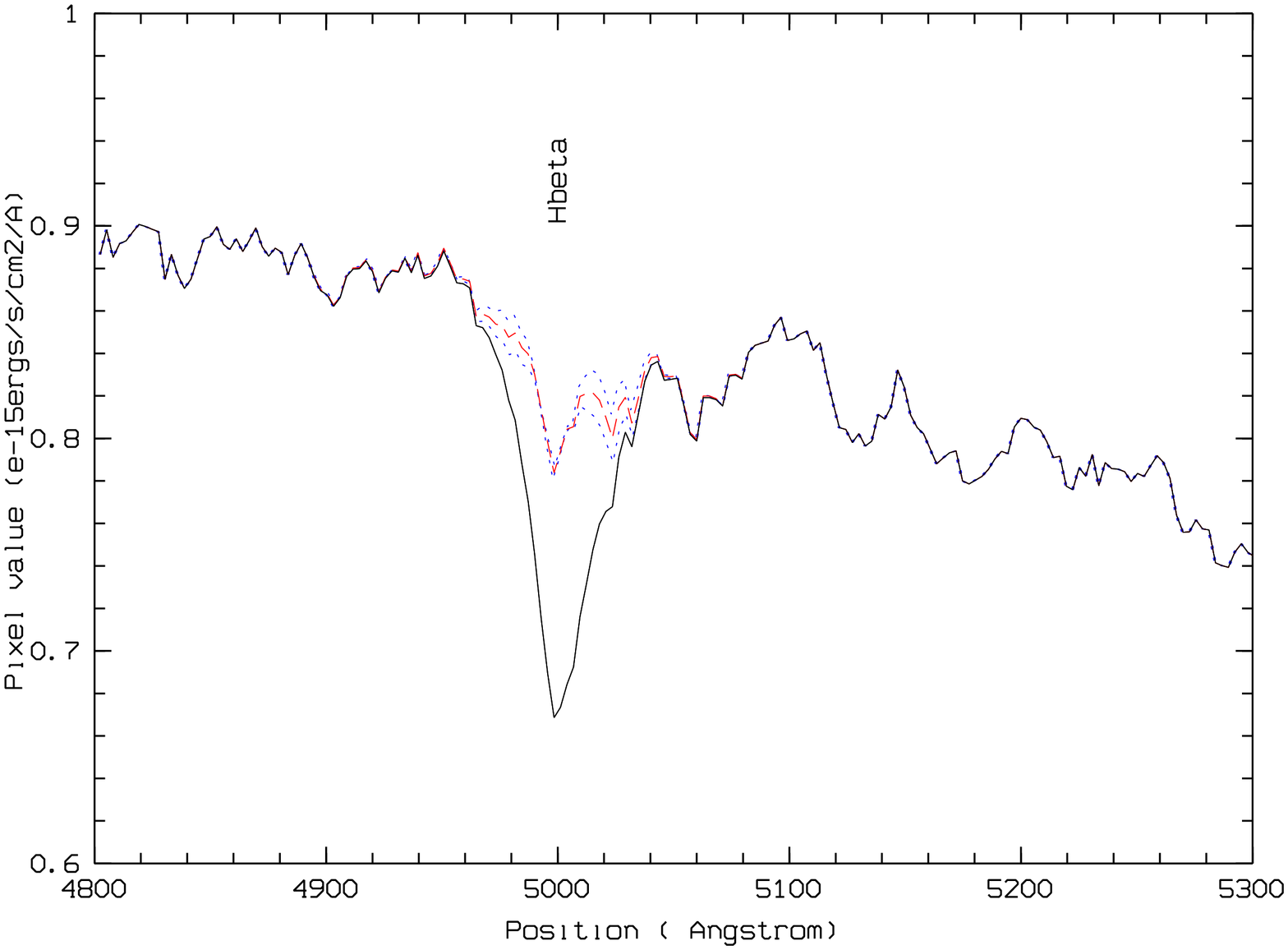}
\includegraphics[width=5.5cm,angle=270]{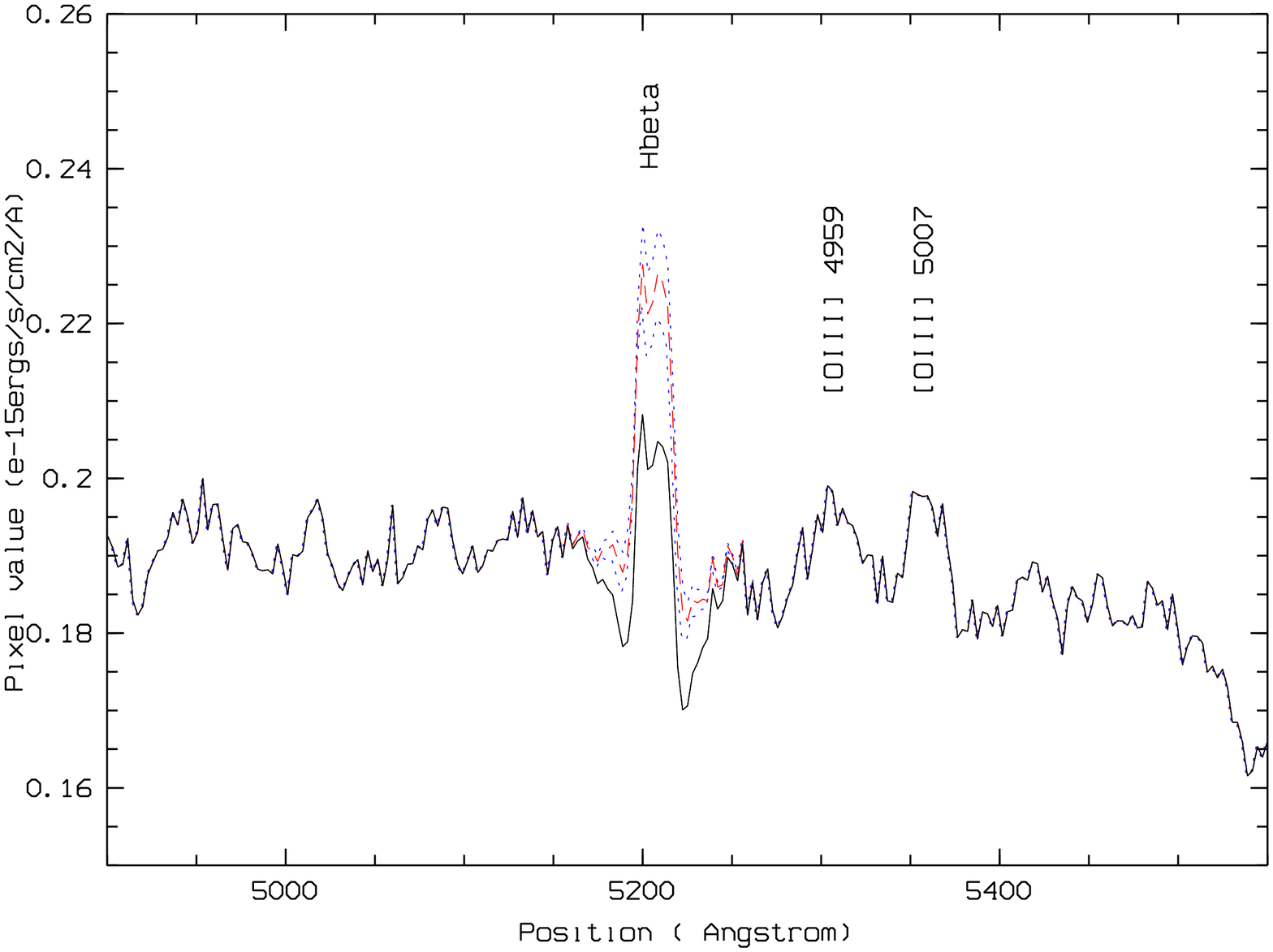}
\includegraphics[width=5.5cm,angle=270]{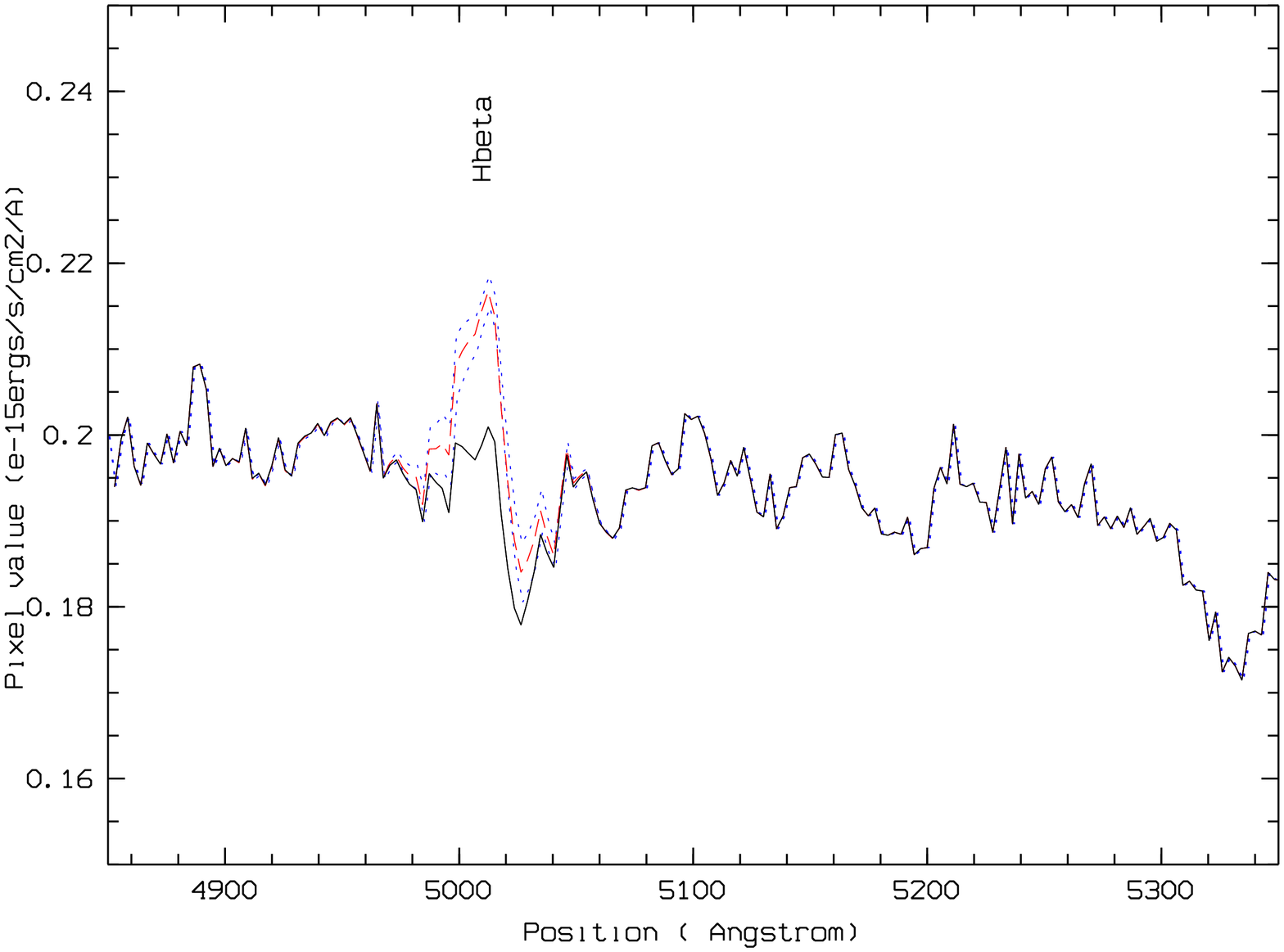}
\caption{Correction for stellar absorption. First panel: test,
correction of a pure absorption H$\beta$ feature. The solid line is
the flux before correction, the dotted ones are the fluxes corrected
by using either H$\delta$ or H$\gamma$, the dashed line is the average of both
corrections. Either of the lines used for the correction 
 gives similar results and the residuals are at the level of the noise. 
 Second panel: application to a blended line (removal of the
absorption line under H$\beta$ emission). Third panel: an emission line
that was too low to be distinguished from the noise 
(because of the absorption by stars) 
 appears after the correction.\label{abs}}
\end{figure}
The Balmer emission lines have a particular importance in the
spectrum: the  ``Balmer decrement'' is  the main 
indicator of dust reddening (see next section); the diagnostic
diagrams  use H$\beta$ and/or H$\alpha$ with  forbidden 
lines to assess the nature of the ionizing source; 
 H$\alpha$ is one of the best star formation indicators. Yet,
atmospheres of A and B stars, generally present in the underlying stellar 
populations of galaxies, absorb in the same lines, thus altering
the resulting spectrum  significantly. 
It is then of particular importance to
correct for this underlying absorption by stars. Several solutions have
been applied in the past, depending on the  quality of the spectra 
and the available information: \\ 
- A constant correction applied to the equivalent width of the emission line. 
The value applied is  usually  around 2\AA\ for HII regions,
and can go up to 5\AA\ (Kennicutt, 1992). This is an average value,  not well 
calibrated on galactic objects. This type of correction is usually
used for  large samples where individual properties are 
neglected and when no other solution is available.\\
- Fit the wings of the absorption lines. These wings  are 
usually seen on both sides of the emission line. This technique has been 
used for instance by Sullivan et al.( 2000)
or Kim et al.( 1995)  to reconstruct both the absorption and the emission 
features. However, this method requires a very good spectral resolution and is 
very sensitive to noise. Fitting a deep absorption line on very faint wings can 
 lead to large errors in the final flux determination. 
 Furthermore, if other emission lines lie close to the Balmer line (like 
for instance [NII]$_{6548}$ and [NII]$_{6583}$) the absorption wings 
will be completely lost, reducing the accuracy of the method. \\
- Use  of the higher order   Balmer lines. This is the  most efficient method, 
which makes use of the  
H$\delta$ and H$\gamma$ lines: the emission decreases rapidly with the order of 
the line whereas the absorption increases. 
One  can compute (e.g.  
Osterbrock, 1989) that the the ratio of the  H$\alpha$ and H$\gamma$ 
emission lines is (case B approximation): 
:$$ E(H\alpha)/E(H\gamma) = 6.1$$ while  from Menzel et 
al. ( 1969), we get, for the ratio of the EW of the absorption lines of the same 
transitions: $$ A(H\gamma)/A(H\alpha) = 1.26$$
We thus have: 

$$ \frac{Abs}{Em}(H\gamma) = 7.7 \times  \frac{Abs}{Em}(H\alpha)$$ \\
This means, as is well known, that  most frequently emission is dominant for the H$\alpha$ 
line whereas absorption is dominant for the H$\gamma$ line.

In the FSM field we have enough spectral resolution and wavelength coverage 
to measure Balmer absorption lines of higher order in all the spectra. In 
the IRAS sample on the contrary, due to a slightly different instrumental 
setting (further red), for many spectra only  H$\alpha$ 
is in the observed range: for those spectra we 
therefore applied the average 
global correction obtained for  the sample (see below).

In practice, to correct the H$\beta$ emission line from underlying absorption, we 
fitted the H$\delta$ and H$\gamma$ absorption lines and derived  
from them the absorption lines at 4961\AA\ and 6563\AA\ 
respectively. The original spectrum is then divided by this absorption 
feature, reconstructing the ``original'' emission line. In Fig. \ref{abs} we
show the efficiency of this method: the first image shows the results 
of the method on a pure absorption spectrum, and the two others 
illustrate how the  H$\beta$ line is recovered, even when it was too low 
(due to the presence of stellar  absorption ) to be distinguished from the noise.

H$\gamma$ is usually blended with a large Mg absorption band, so we 
prefer to use H$\delta$ when available. When both lines are clearly 
detected, we check that the results are consistent (given the error 
bars) and correct H$\alpha$ and H$\beta$ from the average result. 

If none of these two  lines is detectable in absorption, (or if they are in 
emission only), we consider that the absorption by stars is negligible and make 
no correction. In several spectra, the correction makes H$\beta$ pop 
out of the noise. When H$\beta$ doesn't appear even after the 
correction (whereas H$\alpha$ is visible in emission) we deduce that 
H$\beta$ is completely absorbed by dust, and we then use the noise in the continuum  
to estimate an upper value for H$\beta$. 

We find an average correction of 3 \AA\ for the FSM sample and 3.35 
\AA\ for the IRAS sample (average over the corrected spectra -not 
including the non-corrected ones). 

\subsection{Extinction} 
For the FSM sample, we can neglect the foreground galactic 
extinction  because of the high galactic  latitude of the field ($b=-52 \char23$).
 In most FSM galaxies, both H$\alpha$ and H$\beta$ 
are available in emission, so we can use the Balmer decrement to estimate the 
extinction. This method relies on several assumptions: \\ 
- the intrinsic  $(H\alpha / H\beta)_i$ ratio has to be known. Indeed
theoretical
calculations have shown that for a case B recombination this ratio is 
almost independent of density and temperature. We will use $(H\alpha
/ H\beta)_i = 2.86$, the value given by Osterbrock (1989) for a density 
of 100 cm$^{-1}$ and a temperature of 10 000K.\\ 
- the general form of the extinction curve with wavelength is supposed to 
be  known. We used 
the Howarth (1983) extension of Seaton's (1979) extinction law for the 
Milky Way. Other extinction curves 
 used in the literature do not differ widely, at least for the optical range of the 
spectrum.\\ 
We can thus compute the amount of absorption in the visible, $A_V$, for 
each object, using $R=3.2$ (Seaton, 1979) and $X(\lambda)$ as 
described in Howarth, 1983. 
The $A_V$ value is included in Table \ref{results_M}.  \\

\begin{table}[t]
 \begin{tabular}{|lcc|}
   \hline
   sample name & FSM & IRAS \\ 
	&&\\
   number of IR sources:&&\\ 
   - initial IR photometric sample&24&38\\	
   - optical identifications and redshifts&26&28\\	 	 
   - with $H\alpha$ emission line&23&21\\	  
   \hline
 \end{tabular} 
 \caption{ Summary of the two observational samples. Comments: three 
    IR  sources (two in the FSM sample and one in 
   the IRAS sample) have been split into two optical objects.\label{sum-up} } 
\end{table}

\section{Results} 
 
 We try here to characterize the galaxies selected by the 
$170\mu m$ survey, and compare them to the  IRAS sample. 
  All luminosities were computed with a standard cosmology: $\Omega_\Lambda = 
0.7$, $\Omega_M = 0.3$ 
  and we took a Hubble constant of $H_0 = 65$ km/s/Mpc. 
Due to the small distance of the detected objects, the derived parameters  
are rather insensitive to the adopted cosmology . \\
The results of the spectral analysis are presented in Table \ref{results_M} 
for the main emission lines used in the discussion. 

\subsection{Amount of dust} 
\begin{figure} 
\includegraphics[width=5.5cm,angle=270]{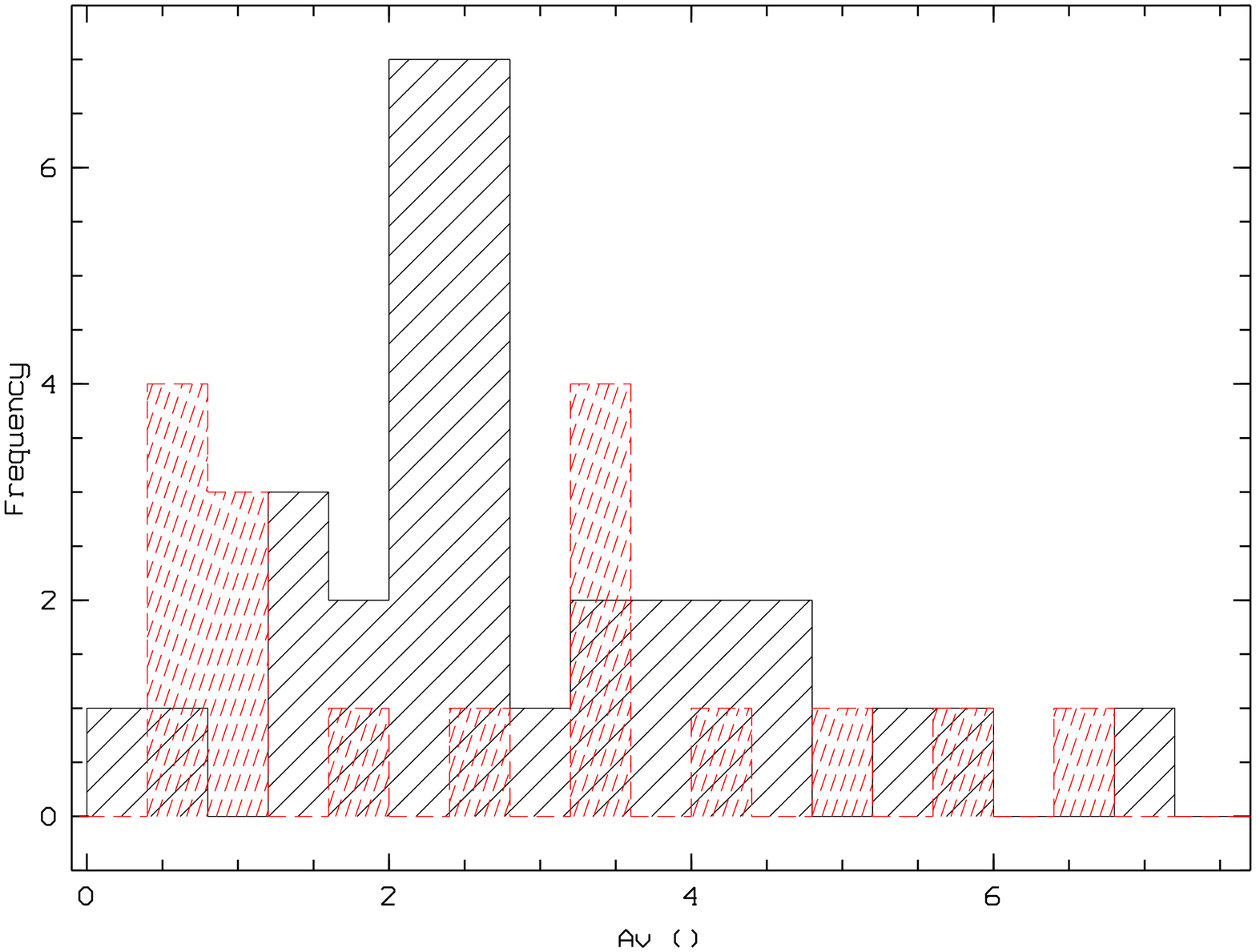} 
\includegraphics[width=5.5cm,angle=270]{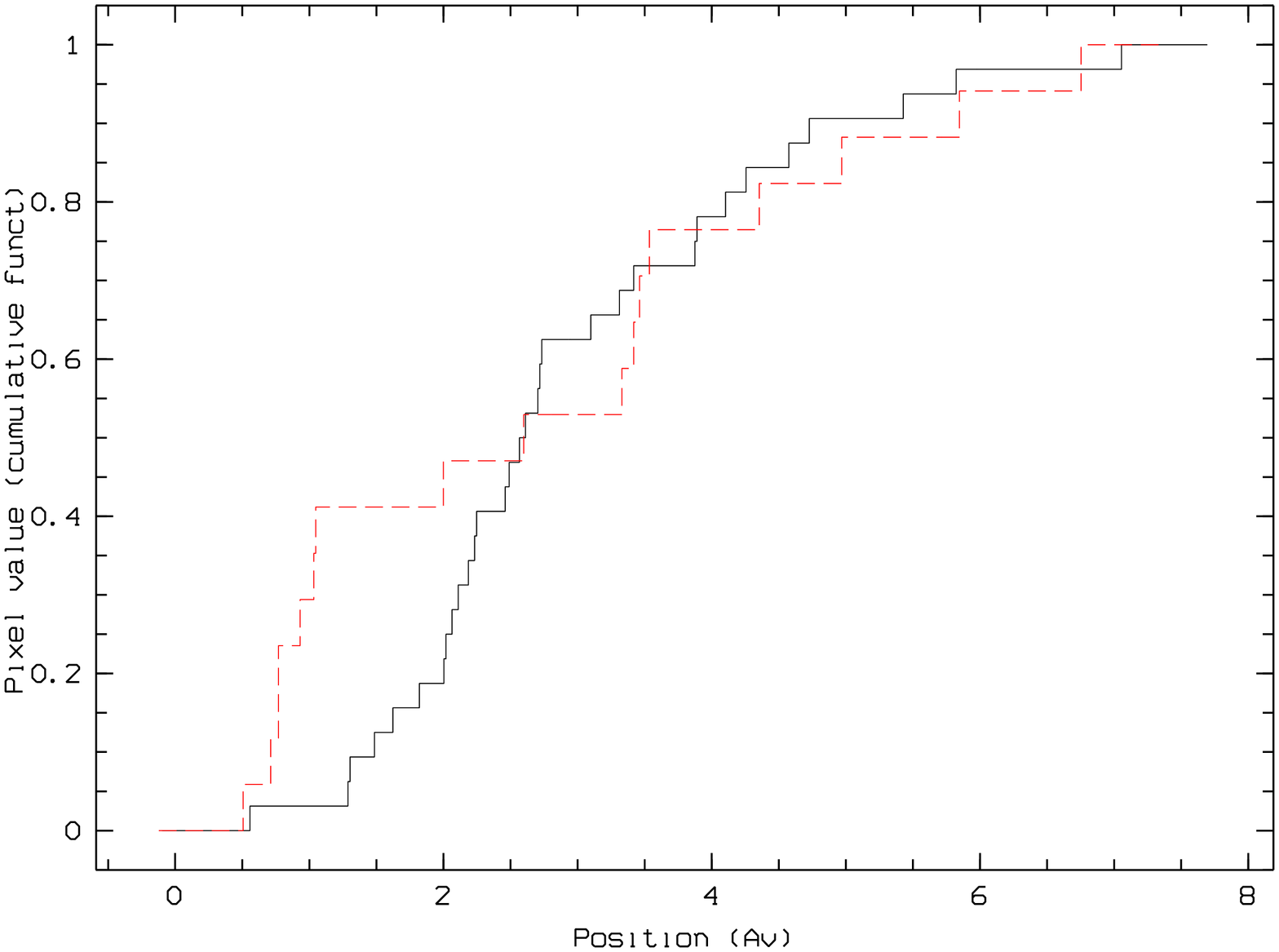} 
\caption{Distribution of absorption by dust. Top: histograms of the 
Av values,  with median hashing corresponding to FSM galaxies 
and inclined hashing to the IRAS sample. Bottom: cumulative function  
of Av for: i) FSM sample (solid curve) and ii) IRAS reference 
sample (dotted curve)\label{Av}} 
\end{figure} 

Fig. \ref{Av} presents the histograms of $A_V$ (after correction from 
underlying stellar absorption) for both samples, and 
Table \ref{list_Av} presents the statistics of the distribution for all 
 spectra observed (excluding  occasional stars). The reason to use for 
the  statistics  all the observed spectra 
 is because when several galaxies are associated to one IR 
source, an ``average'' extinction factor would have no real meaning. 
 The result of a Kolmogorov-Smirnov test 
on the cumulative functions shows that the two samples have a 
reasonable chance of coming from the same initial distribution (a $D$ 
parameter of less than 0.4,  with $n=17$, leads to a $p(D)=25\%$, that is 
to say that two samples drawn from the same distribution have 25\% 
chances of being farther apart than these two samples; see Miller, 
1956). 

We conclude that the two samples are remarquably alike as far as 
 extinction is concerned ($A_V\sim 3$ for both  samples). 
They are also very similar to the ``Bright galaxies sample''(Veilleux 
 et al. 1995, thereafter BGS), where   $A_V$ has an average value of 3.6 ( with our adopted value of  R=3.2), but rather far from  
UV-selected samples ($A_V\simeq1$; Sullivan et al, 2000 for the 
FOCA galaxies). The extinction is also  significantly higher than in 
the CFRS $z<0.3$ sample, where 
 $A_V=1.52$ for objects selected in the I band ( 
and is an upper limit,  as no correction has been applied there  
 for underlying 
stellar absorption). 
 This effect is not surprising: it is 
quite clear  that $170\mu$m and $60\mu$m observations both select 
heavily obscured objects. 

\begin{table}[b] 
 \begin{tabular}{llllll}
   Sample&Nb&$<A_V>$&$A_V$min&$A_V$max&disp\\
   name&of spect.&&&&\\
   \hline\\
   FSM&33&2.88&0.18&7.1&1.5\\
   IRAS&17&2.71&0.51&6.8&2\\
   \\	
 \end{tabular}
 \caption{Extinction parameter for individual galaxies in both samples. 
The two distributions 
   are similar, given the small statistics,
 and cover the same range.\label{list_Av}} 
\end{table}

 We want to stress  that an extinction  of $A_V = 2.9$ magnitudes 
 (average value) results in a reduction of a factor 9 (or an optical 
 depth of $\tau(H\alpha) = 2$) in the  H$\alpha$ line intensity, and a 
 factor of 50 ($\tau([OII]) = 4$) for the [OII] blend, as compared to 
 a factor of 2 in H$\alpha$  and 4 in  [OII] when we take  the usual 
 prescription of $A_V = 1$ only.  It is thus important to characterize 
 the type of  object selected , in order to apply the appropriate 
 correction.  

\subsection{Redshift distribution}
\begin{figure}
 \includegraphics[width=5.5cm,angle=270]{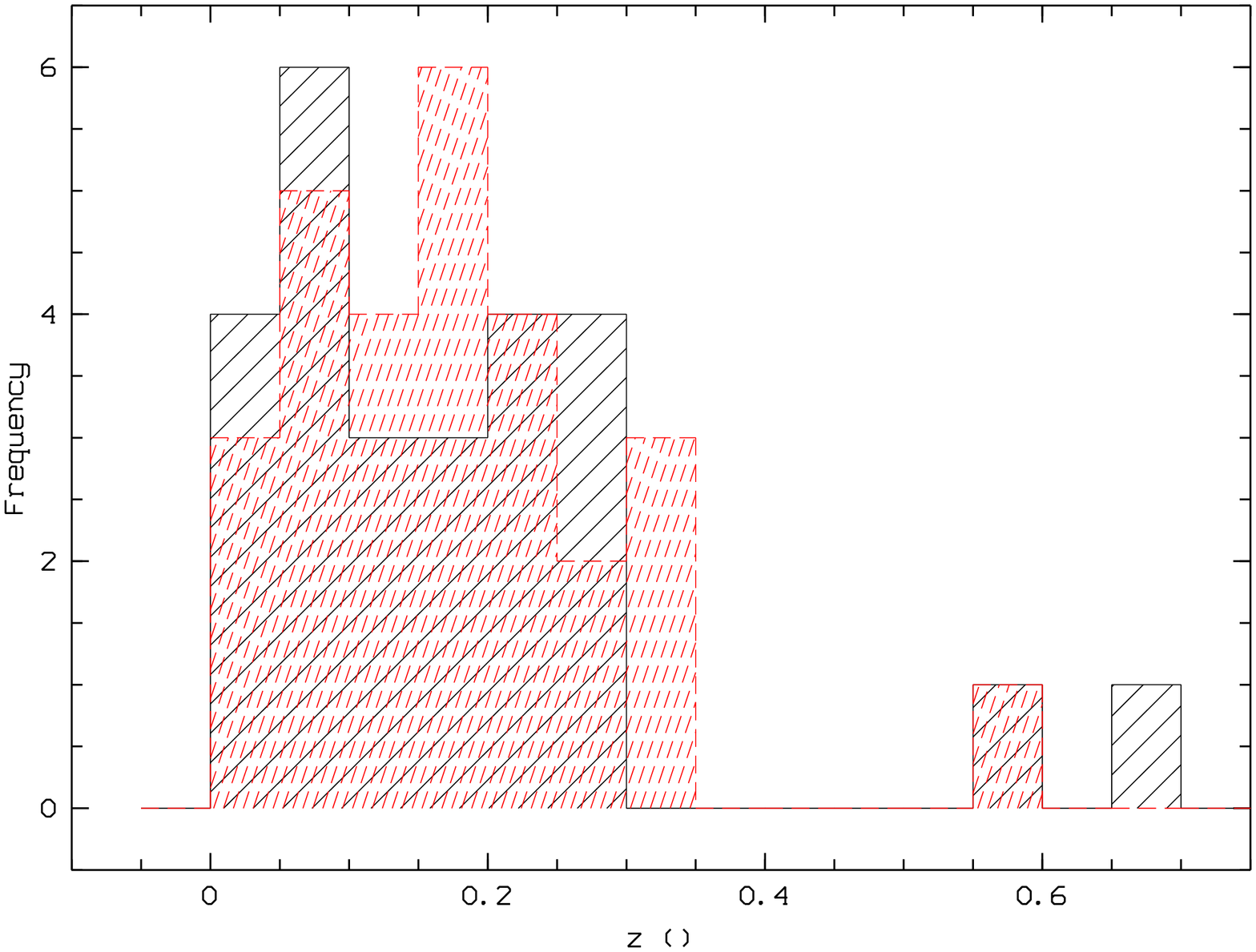}
 \includegraphics[width=5.5cm,angle=270]{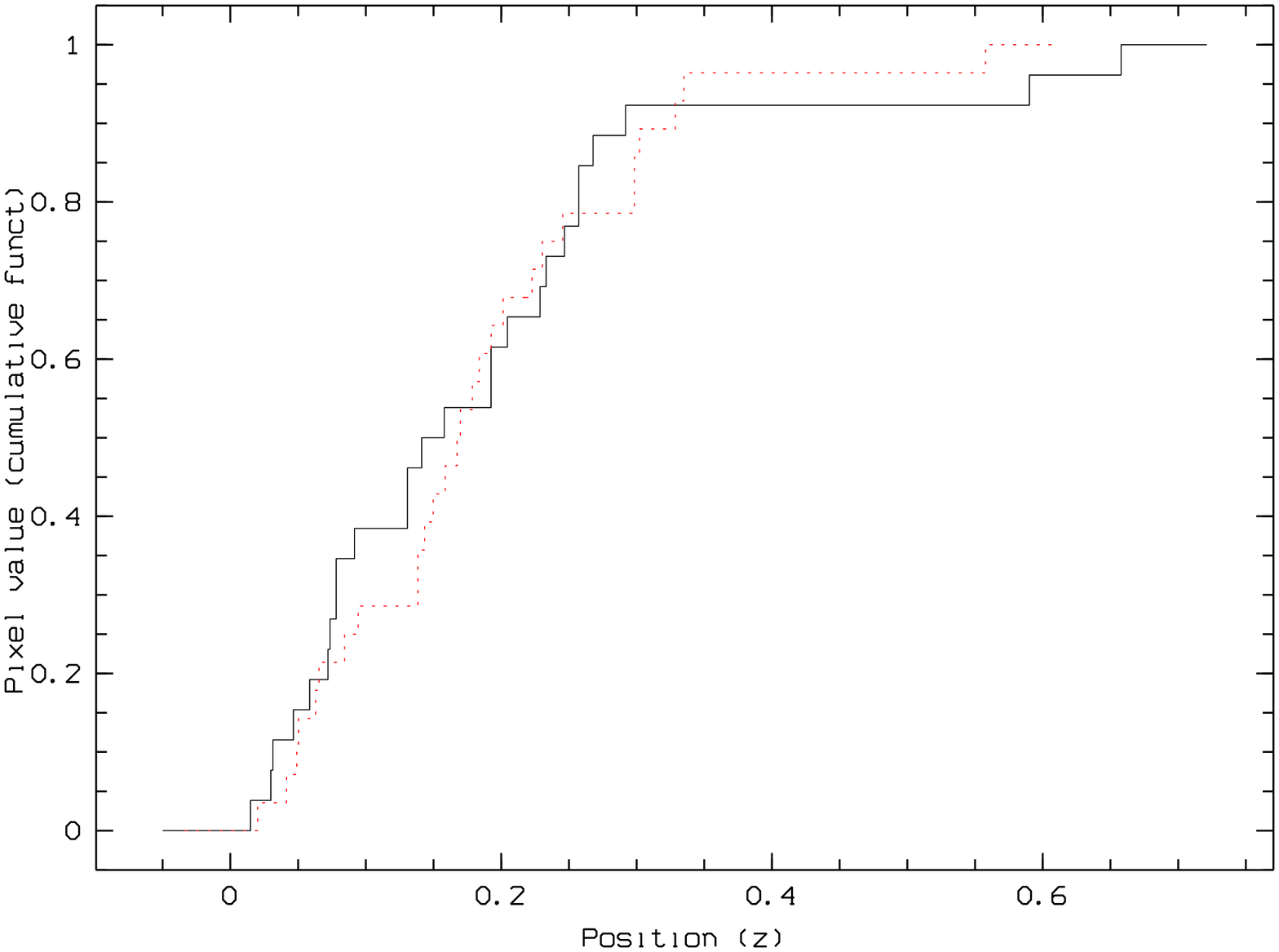}
 \caption{Distribution of the redshifts of the IR sources.\label{z}}
\end{figure}
We show in Fig. \ref{z} and Table \ref{list_z} the statistics of the 
obtained redshift distributions.  It is quite obvious that the 
repartitions are similar. The Kolmogorov Smirnov test tells us that two 
stochastic samples taken from the same parent distribution have a 
probability of more than 70\% to be more different than these two. 

The only two objects at larger redshift ($z>0.3$) in the FSM  
sample are Seyfert galaxies. The brighter sources in the  
 170 $\mu$m selection are thus rather close by. \\ 
\begin{table}
 \begin{tabular}{llllll}
   Sample&Nb&$<z>$&$z$ min&$z$ max&$\sigma$\\
   name&of spect.&&&&\\
   \hline\\
   FSM&26&0.18&0.015&0.658&0.16\\
   IRAS&28&0.18&0.020&0.558&0.12\\
   \\	
 \end{tabular}
 \caption{Statistics on z for both samples. The two distributions 
   are obviously very alike.\label{list_z}} 
\end{table}

\subsection{Emission lines and optical properties} 
 We will use classical diagnostic 
diagrams to derive the nature of those galaxies.   

In the FSM sample, we usually have secure measurement for (at least) the 
following lines: [OII] (3727\AA), H$\beta $, [OIII] (4959\AA\ , 
5007\AA), H$\alpha$, [NII] (6548\AA\, 6583\AA).   
There is no broad emission  
line object in the FSM sample. Although our statistics is small (only 20 
objects in the brighter part studied here), this is consistent with the 
results of the BGS where   Veilleux et al., 1995 found only two
Seyfert 1 galaxies out of the 140 spectra analyzed, and they  are very
bright IR objects.\\  
The fraction of emission-line galaxies in our sample is very 
high, confirming the pertinence of an IR-selected 
sample to find star forming objects. To all but one (FSM\_18) 170~$\mu$m 
sources, the optical corresponding object is one (or several)
emission-line galaxy.
Except for this single case, all spectra can be classified ``emission + 
absorption'' (all spectra have Balmer 
emission, most have Balmer absorption and metallic absorption lines), 
see Fig. \ref{spectres}. 

\begin{figure}[t] 
\hspace{-0.4cm} 
\includegraphics[width=6.5cm,angle=270]{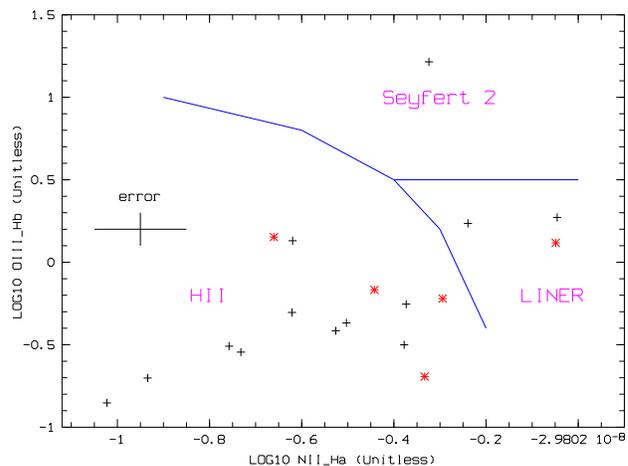} 
\caption{Diagnostic diagram for both our samples. + : FSM sample, * 
: IRAS reference sample. All emission-line measurements are corrected 
for dust extinction and absorption lines. Only spectra with H$\beta$,
{[}OIII], H$\alpha$ and {[}NII] can be plotted: due to a smaller
spectroscopic range, very few IRAS objects could be plotted (see
text). The cross at the left indicate the average error bar. \label{diagn}} 
\end{figure}

Fig. \ref{diagn} presents one of the most classical diagnostic 
diagrams (Baldwin et al., 1981) to separate HII region-type 
galaxies (bottom left of the diagram) from Seyfert 
2 (upper right of the diagram) and LINER (Low 
Ionisation Nuclear Emission Line Regions, bottom right). The frontiers 
are from Dessauges-Zavadsky et al. (2000), but  a variety of empirical 
separations can be found in the literature, and the  
 definition of the regions is rather approximative. 

 This diagram  is well  adapted to our data. The 
signal to noise ratio of our spectra does not allow us to use [OI] or [NeIII] 
lines (usually not detected), or the  [SII] often affected by atmospheric features. 
 For  objects at 
$z>0.3$, $H\alpha$ is out of the observed range, so we  use also the  ``R23'' 
criterion from Dessauges-Zavadsky et al. (2000): 
$$ R23 = \frac{[OII]_{3727}+[OIII]_{4959} + [OIII]_{5007}}{H\beta} 
$$$$ 
{\rm and \:\:}R23>1.1\:\: {\rm for\: Seyfert\: 2\: objects.} 
$$

One object is classified ``Seyfert 2'' by both criteria: FSM\_004. 
The  ``R23'' test  points out to two more objects, FSM\_011 and FSM\_035, 
for which $z>0.5$ and thus we don't have the H$\alpha$ line. These latter two  
 objects  are quasi-stellar (punctual 
appearance on acquisition image) and at intermediate redshift 
($z\simeq0.6$); they are the most remote  objects of our sample. 
The excitation of FSM\_11 (and absence of broad H$\beta$) indicates 
a Seyfert 2 galaxy, while  FSM\_35/1  is clearly a QSO. 
FSM\_4 is a 
nearby active galaxy at $z=0.0784$. 

Two FSM objects are classified as ``LINER'' on the diagram, FSM\_013 
and FSM\_003 (though this last one is not secure, given the error bars). 
Unfortunately, the classical test for LINER is not applicable here 
since the  [OI] 6300\AA\ line has  not been detected  in our 
spectra. 

 We thus get  3 to  5 active objects out of 29 in the FSM  
sample, that is  11\% to 19\% of the objects. The IRAS reference 
sample shows similar results, though the statistic is much 
smaller due to the smaller observed range and spectral resolution.  

These values are similar to the CFRS results (between 8\% 
and 18\% when  including  the effects of stellar absorption), but 
rather far from the  high number of AGN found in the IRAS bright 
galaxy sample. If this effect is confirmed (by larger  
samples) it tends to show that the faint IR galaxies are indeed a 
 population different from the bright IR objects: heavily obscured, but 
with a majority of star forming objects and only few AGN. This is consistent 
with the finding (e.g. Lutz et al., 1998; Veilleux et al. 1999) that 
the fraction of AGN increases strongly when going to   ULIRG objects. 
Indeed, as shown in Section 5.5, the IR luminosities  of the bright FSM objects 
are rather modest ($\rm  L_{IR} < 10^{12} L_{\odot}$) and, at those 
luminosities,  the fraction of 
AGN's is smaller than 20\%  in the 1 Jy sample studied 
by Veilleux et al., for instance.  It  increases abruptly to 50\%  only  at 
$\rm  L_{IR} > 10^{12.3} L_{\odot}$. \\
  
{\bf
\subsection{IR temperatures} 
\begin{figure}[!t] 
\includegraphics[height=12cm, width=9.cm]{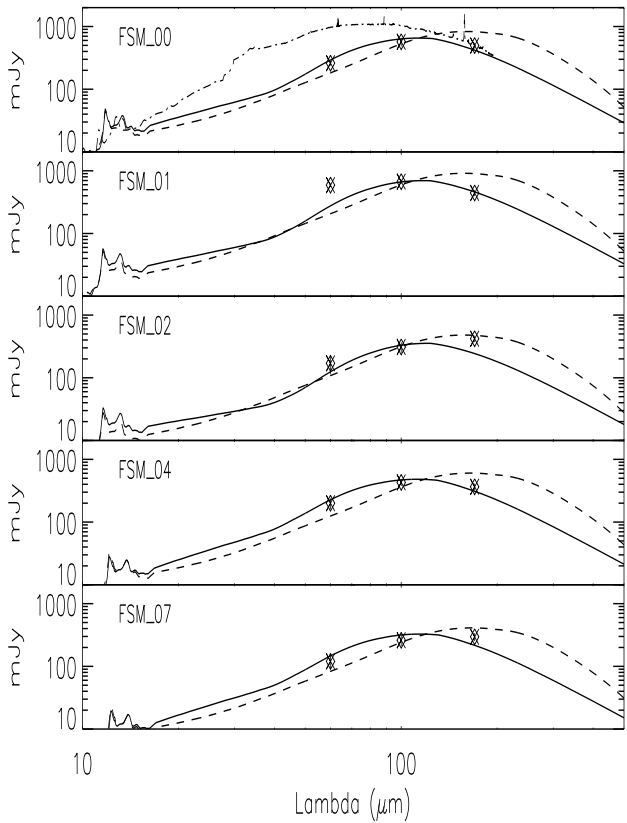}
\caption{SED modelisation compared to five observed 
cases. Solid line: starburst template, dashed line: Cold
template. Both templates are from Lagache et al. (2003)
Also shown for comparison is the spectrum of M82 (dashed-dotted)
\label{model}} 
\end{figure} }

One of the main characteristic of our galaxies is their cool IR 
temperature,  which can be determined when we have more than one 
observational point. As indicated in Table \ref{list_FIRBACK}, for  5
out of 29 objects, we have, in addition to 
the 170~$\mu$m fluxes, also the 60 and 100 $\mu$m IRAS measurements,  
obtained from the  SCANPI procedure. 
For another  13 galaxies, this 
method gives only upper limits for 60 and/or  100$\mu$m fluxes.\\

To characterise the dust emissivity of these galaxies, we studied 
three quantities: the greybody temperature $T$, the infrared colors and 
the spectral energy distribution.

The greybody temperature $T$ was computed for direct comparison with other 
galaxies from the literature. We use a spectral index of $\beta = 
1.5$, and fit a modified blackbody to the infrared measurements at 60 
and 170$\mu$m for  5 galaxies, and use the upper limits at 60$\mu$m for 
 13 other galaxies. We find an average temperature of 22.7K, with values 
ranging from 20.9K (FSM\_002) to 24.7 (FSM\_001) for the first  
 5 galaxies, and a somewhat wider distribution when we work also with  upper 
limits: average temperature of 23.5K with values ranging from 19.5 
(FSM\_005) to 32.7K (FSM\_011). This results indicate a  rather 
cold population.

In addition to the greybody temperature, we computed 
the IR color $\log(S_\nu (100\mu$ m$)/S_\nu (60\mu$m)), as  usually done for IRAS
galaxies. We find a mean ratio of -0.32, with a dispersion of 0.2 
dex. This is highly compatible with the ``normal'' IRAS galaxies: 
Soifer et al.(1989; their Fig.1) show that the histogram of the color 
distribution  of all the BGS galaxies is  centered at -0.3. A similar 
value is obtained  with our own small IRAS sample. 

 Thus the FSM galaxies  have the same average color as normal IRAS 60 
$\mu$m selected galaxies, but the observations at 170$\mu$m  show a 
cooler dust temperature. This is generally the result whenever a longer 
wavelength point is added to IRAS measurements, as a single temperature 
blackbody is probably too crude an approximation to reproduce the SED of 
those objects.

 To investigate further the IR spectral energy distribution  
of these galaxies, we fit our observational data  points
with the ``starburst'' and ``normal'' galaxy  templates from
Lagache et al. (2003). The ``normal'' galaxy  template
is colder than the ``starburst'' template, with a peak
at around 180 $\mu$m. Results of the fit are  shown in Fig.
\ref{model} and detailed below:
\begin{itemize}
\item FSM\_000: the best fit is obtained with the ``starburst'' template with  L$_{FIR}$=10.95
\item FSM\_001: the best fit is obtained with the ``starburst'' template with L$_{FIR}$=10.6. 
 The 
60~$\mu$m point is however not well reproduced: this can be due to the existence of a hotter component.
\item FSM\_002: the best fit is obtained with the ``normal'' template with L$_{FIR}$=10.35.
\item FSM\_004: the best fit is obtained with the ``starburst'' template with L$_{FIR}$=11.25.
\item FSM\_007: Both templates give only an approximate  reproduction of  the measured fluxes.
\end{itemize}
Note that none of these galaxies has a SED which resembles the one of M82. 
Although the sample here is small, it appears that our objects are best reproduced by a 
cold starburst template, giving luminosities typical of LIRGs.

\subsection{IR luminosity}

As for most objects in our sample, we do not have the 
 IRAS (60$\mu $m, 100$\mu $m) flux measurements, 
 we cannot use the classical formula (Helou  
et al., 1988) to derive the total IR luminosity ($\rm L_{IR}$). 
We are thus obliged to use models or template SED to derive it in a 
homogeneous way for all objects in the sample.  

We first performed a series of tests to determine the best way to 
compute the total luminosity. Three different methods are compared   
(we note the disadvantages of each method in italics):\\ 
1 -  take in the grid of spectra proposed by Lagache et al. the 
one that is closest to the observed points. {\em This supposes a 
universal relationship between luminosity and color}. \\ 
2 - use the  normal  template SED from Lagache (built from a number 
of FIRBACK objects).   
 {\em This supposes a universal shape for 
all objects}. \\ 
3 - fit only the shape of the model of Lagache et al., regardless of 
its intrinsic absolute value at peak, and then compute the total 
luminosity. {\em This method is correct only for those objects were we  have 
more than one observational point} (9 for the FSM  
sample, 3 for the IRAS reference sample).\\   

Though the three methods lead to very different colors for a given  object,  
 the computed total luminosities  do not differ by more than 20\% 
(and 10\% on average), which is also the value of the error bars of 
 the ISO and IRAS  data. 
 We conclude that  any of the two first methods can be used to 
derive the total luminosity for all objects in a homogeneous way, with 
a resulting uncertainty of less than 20\%. The third method can be 
 used as a check of 
this uncertainty for the 6 objects where more than one measured point is 
available. Note that this "total" IR luminosity is about 2 times larger than 
the FIR usually derived from the IRAS data alone (40 - 200$\mu$m; 
 Helou et al. 1988). \\ 
$\rm L_{IR}$ has in the end  been determined using the first method, and 
the error is then  of the order of 0.1 dex. 

\begin{figure} 
\includegraphics[width=8.5cm,angle=0]{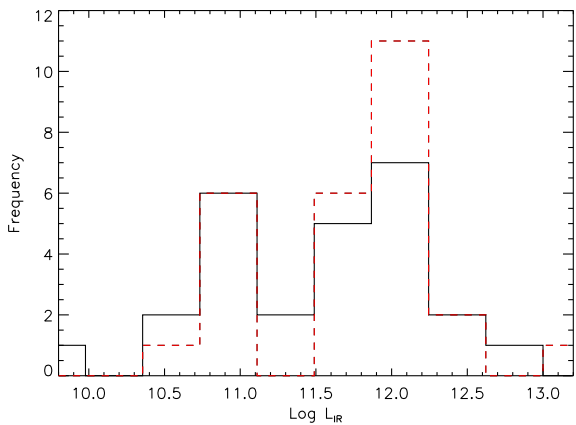}
\caption{Distribution of IR luminosities ( from 0.4 to 1000~$\mu$m)
for  i) the FSM sample (solid histogram) and ii) the IRAS 
sample (dashed histogram).\label{LIR}}  
\end{figure}

Most of the FIR selected galaxies are LIRGs (Luminous Infrared Galaxies) 
with $\rm 10^{11} < L_{IR} < 10^{12}~ L_{\odot}$, and many of them have lower 
luminosities, down to $\rm L_{IR} \simeq 10^{10.5}$. Very few are ULIRGs: 
the only two objects  belonging to this category are the higher 
redshift  AGN's found here in the FSM. 
Fig. \ref{LIR}  shows the distribution of $\rm L_{IR}$ and 
Table \ref{list_LIR} 
gives the statistics on $\rm L_{IR}$. 

We finally derived the IR to optical luminosity ratio as  follows:\\ 
- $\rm L_{IR}$ computed with the first method;\\ 
- optical luminosity computed by  integrating the  continuum from 4000 to 
7000 \AA\ in the rest frame (range directly accessible with our spectra).
We find ratios between 10 and 100, with an average of 65. These are 
rather high values, indicative of high extinction and/or strong activity. 
On the  two higher redshift FIRBACK source identifications 
($z=0.45$ and $z=0.91$) found in FN1, Chapman \& al (2002) derived these ratios 
between 10 and 15; but correcting for the different definition they used
(standard IRAS FIR, and a 4 times broader range for the optical
luminosity), it appears that their values are entirely compatible with
those we find here in this larger sample.  

\begin{table}[h]
 \begin{tabular}{cccccc} 
   Sample&Nb&average&$\log(\rm{LIR})$&$\log(\rm{L_{IR}})$&$\sigma$\\ 
   name&of spect.&$\log(\rm{L_{IR}})$&min&max&(dex)\\
   \hline\\
   FSM&26&11.5&9.6&12.9&0.8\\ 
   IRAS&27&11.7&10.6&13.0&0.6\\
   \\	
 \end{tabular}
 \caption{Statistics on total IR luminosities (from 0.4 to 1000~$\mu 
   $m, in units of solar luminosity) for both 
   samples.\label{list_LIR}}  
\end{table} 

\subsection{Star formation} 
Several indicators are generally used to determine the star formation rate of 
galaxies, the two most  useful ones in  our sample being the  H$\alpha$ emission line and the IR luminosity.  
 
The  H$\alpha$ emission line fluxes used to derive   the star formation rate 
are corrected 
for extinction: this is indispensable as our objects are heavily obscured.  
For the conversion factor, we use the  values from Kennicutt (1983):  

$$ \rm{SFR\:(M}_\odot\rm{per\:year)} = 
\frac{\rm{L}_{H\alpha}\:(10^{41}\rm{erg/s})}{1.12}$$ 
 
To compute the star formation rate from the IR luminosity, we 
use the calibration of Devriendt et al. (1999)  which is  consistent with 
the SED models  used:
 
$$ \rm{SFR\:(M}_\odot\rm{per\:year)} = 
\frac{\rm{L}_{IR}\:(10^{9}\rm{L_\odot})}{7.7}$$   

The results for both samples are presented in Table 
\ref{list_SFR}. The correction applied to the  H$\alpha$ flux  seems 
to be insufficient, as the SFR is lower than the one   calculated 
from the IR luminosity: the  star forming regions are probably completely 
screened in the optical, as suggested by the rather high values of
optical depths derived for H$\alpha$ in Sect. 4.1.   
However, the objects we observed do not  have large   Star Formation 
Rates: with the exception of one or two more "extreme" objects, 
the  average SFR is a few tens of  solar masses per year, which corresponds   
  to moderate  starburst episodes. 

\begin{table}

 \begin{tabular}{lcccc}

   \multicolumn{5}{l}{\bf Star formation rates ($M_\odot /$yr) from 
   corrected H$\alpha$ flux:}\\
   &sample size:&average:&min:&max:\\
   FSM&20&9.7&0.3&39\\
   IRAS&20&37&0.5&440\\
   &&&&\\
   \multicolumn{5}{l}{\bf Star formation rate ($M_\odot /$yr) from IR luminosity:}\\	
   &sample size:&average:&min:&max:\\
   FSM&21&41&0.5&990\\
   \hspace{0.2cm} (with H$\alpha$)&20&41&0.5&320\\	
   IRAS&27&65&2.6&1300\\
   \hspace{0.2cm} (with H$\alpha$)&20&106&5.2&340\\	
   &&&&\\
 \end{tabular}
 \caption{Distribution  of the star formation rate for the FSM and 
   the IRAS sample (  computed from the corrected H$\alpha$ emission 
   flux and from the total IR luminosity;  
    only  objects classified as ``HII region-like''  in section 
   4.3 have been used). For comparison, the SFR inferred from the IR 
   is also shown for 
   the sub-samples with a measurable H$\alpha$ line. 
   }
 \label{list_SFR}
\end{table}

\section{Discussion on the nature of the galaxies}

 We observed the brightest objects in the  FIRBACK South Marano sample, 
and were almost 
complete in our follow-up of the brightest sources (20 out of 22 IR 
sources in the 4 $\sigma$ catalog). Pending the follow-up of 
the faintest objects in the FSM complementary catalog (15 
objects with $ 3 <\sigma < 4$), and the study of similar sources in the 
northern FIRBACK fields,  we have already here with this sample a 
fairly good picture of the nature of the brightest 170 $\mu$m  sources.

Because of the so-called "negative K-correction effect" (e.g. Dole 
et al. (2001)), a deep flux limited survey in a far-IR band like 
FIRBACK is expected to be biased towards  distant galaxies. 
Our results however point to {\em a population of galaxies much nearer 
than  expected}. Sajina et al. (2003) described the 
$T/(1 + z)$ degeneracy in imaging infrared surveys such as 
FIRBACK. The results of our follow-up shows that the 170 
$\mu$m luminosity is more an effect of low temperature than one of high redshift, at
least for the brightest objects. This has also been seen by 
Chapman et al. (2002), who, upon selecting just  two FIRBACK objects  
supposed to be at very
high redshift,  found in fact redshifts much lower than expected. 
We show that this property is more general and that we systematically 
overestimated the sources's redshifts. 
 
 The physical properties of the galaxies in this 
sample can be summarized as follows. 

First, as expected, they are heavily extincted. Measurements of 
$A_V$ give an average value of about 3, with extreme cases up to 7, 
 consistent with classical IRAS galaxies (Sanders \& Mirabel, 1996) . 
 A very high proportion (more than 80\%) of the galaxies shows narrow  
emission lines. The   origin of the IR emission at 
170$\mu m$  is therefore likely to be a central/circumnuclear engine heating 
the  dust, just as  in IRAS galaxies. The total infrared luminosity 
produced by this source is moderate in both our IRAS faint galaxies and in the 
FIRBACK galaxies: they can be classified as ``luminous infrared 
galaxies'', with $L_{IR} \simeq 10^{11} L_\odot$, but not ULIRG's. 
 The dominant energy source  is   star formation in most 
cases ( $\simeq 85 \%$ of the galaxies), as indicated by the analysis 
of the emission spectrum.  
 Among the 15\% of objects possibly dominated by an AGN, two are the 
most luminous and farthest objects of the sample;  only one is 
dominated by an active nucleus in the $4\sigma$ sample consisting of 
nearby, bright objects.  
 Consistently with the properties of IRAS galaxies, we find that: i) nearby 
luminous infrared galaxies are mainly powered by star formation 
and ii) the proportion of AGN increases with the IR luminosity.

 The homogeneity of the properties of 80\% of the FIRBACK galaxies 
analyzed so far allows us to consider them  as a population in itself, 
characterized by 
the following properties: a small redshift ($z\simeq0.1$), a moderate 
IR luminosity ( $L\simeq10^{11}L_\odot$), a  peak 
of infrared emission between 100 and 200 $\mu$m, and a main energy source due 
to star formation with moderate SF rates ($SFR \simeq 10 M_\odot/yr$). Is this 
population ``new'' or is it part  of another, already known, class of 
galaxies? To answer this question,  we can  compare this FSM sample 
to three other samples of infrared galaxies: the 15$\mu$m-selected sample 
of Elbaz et al. (2002), our sample of faint IRAS galaxies, and the 
classical IRAS galaxies in  the BGS ( Soifer et al. (1987), Kim et  
al. (1995), Veilleux et al. (1995), under others). 

The comparison with the galaxies of Elbaz et al. (2002) in the ISOCAM 
15 $\mu$m survey shows that the 
two population differ rather widely. Elbaz et al. found a 
higher density of objects in the HDF-N (40 galaxies at $S_{15} > 
0.1$ mJy over a field of $26'^2$), but their survey is also deeper:  
0.1 mJy at $\lambda = 15 \mu$m corresponds to 4.5 mJy at 
$\lambda = 170 \mu$m with our template SED, and this  leads to galaxies  
more luminous on the average, and  at higher redshift ($<z> = 0.8$). 
 But the FSM galaxies cannot be simply interpreted as corresponding to the 
closest and  coldest tail of the distribution of  Elbaz's  objects: 
 those  have a very low 
infrared luminosity and usually no emission line, contrary to the FSM galaxies. 
The FSM galaxies resemble in fact much more to their 
 highest redshift objects. The two populations are therefore quite different. 
The comparison can  however not be pushed much further, in particular 
the comparison of the SED's,  as no far-IR measurements are available 
for Elbaz's sample.

On the contrary, the comparison of the FSM galaxies with the very faint IRAS 
galaxies selected at  60$\mu$m 
 shows that the two populations are very similar. All the 
properties we were able to measure on both samples are close to identical: 
$z$ distribution, average IR luminosity, nature of the 
powering source, and star formation rates. { \em With the 
available  data, it is impossible to distinguish statistically  
between the two samples}. Unfortunately, because this IRAS sample corresponds 
to the faintest detectable galaxies (Bertin et al. 1997) at $60 \mu$m, 
which is the most sensitive band in the IRAS survey, no $100 \mu$m data 
are available for them and therefore no dust temperature 
estimate can be made, nor a SED derived.  
Further measurements with SIRTF will be desirable to better characterize 
 those objects.   

If  we now  compare the FSM galaxies  with the more classical IRAS galaxies 
like  the BGS for example, we find many similarities (spectral properties, IR 
luminosities) but one significant difference:  the shape of the infrared SED. 
 Other deep 
samples of IRAS galaxies have focused on the search for ULIRG's and therefore 
selected, by definition, objects much warmer than the FSM galaxies. It is 
therefore 
quite plausible that the majority of faint, unstudied, IRAS galaxies have on 
average a colder IR emission than assumed up to now (and than measured in 
the few 
$60 \mu$m selected samples). 
Although we cannot exclude at this stage of analysis that the 
galaxies picked up here at $170 \mu$m and the very faint IRAS galaxies 
have in fact an even  colder infrared emission than 
the main IRAS sample, it has been shown in the ISOPHOT serendipity sample by 
Stickel et al. (2000) that a cold dust component was present in many 
galaxies, so that all these objects could be similar, but 
only  a detailed analysis  of SED's will definitely answer this question. 
 \\

\section{Conclusion} 
 The brightest objects of the FIRBACK South Marano survey sample 
the same population as the IRAS faint survey: a  numerous population of cold,  
dusty, nearby galaxies. These galaxies are characterized by:\\ 
i) a high density of sources (about ~20 per square degree) in  
a small volume of the universe, since all these sources lie at $z\simeq0.2$. \\
ii) a cold IR emission, with a peak of the emission closer to $170 \mu$m 
than to $100 \mu$m.\\ 
iii) a moderate IR luminosity in the LIRG range 
($10^{10.5} < L_{IR} < 10^{12}$), quite similar to the 
``average'' IRAS galaxies\\ 
iv) a high proportion of star forming objects, though with moderate 
SF rates (around ten solar masses per year), and a very small fraction
of active galactic nuclei. 

The importance of this "quiet" population of IR galaxies was clearly
underestimated in the analysis of the IRAS survey and has to  be taken
into account in the calculations of the far-IR background. It is now
clear that the far-IR background cannot be reproduced only by high-z
galaxies (Lagache et al. 2003), but that closer and colder galaxies
are contributing  also. Even if their contribution is probably rather
low, they will form an important foreground for future
infrared deep field observations.
The relative contributions of nearby galaxies and of distant objects in the 
far-IR selected samples in particular, and in the far-IR background in
general, will be estimated more precisely when we have completed the
survey of fainter objects in the FSM and in the northern fields of the
FIRBACK survey.

\begin{figure*} 
\caption{Identification charts for FIRBACK South Marano field. DSS
optical image superimposed to the error circle of infrared ISO
source and radio emission (black mark). The straight line show the
position of the slit (1.2'' large and 5' long), and an arrow indicate
a galaxy with an available spectrum. The number associated to each
arrow corresponds to the number in table \ref{results_M}, with an
asterisk indicating that the galaxy is associated to the infrared
emission. \label{cartes}}
\end{figure*}

\begin{figure*} 
\caption{ Spectra}
\end{figure*}

\end{document}